# Raman spectroscopy of ion irradiated SiC: chemical defects, strain, annealing, and oxidation

Alexander J. Leide[a1*], Matthew J. Lloyd[a], Richard I. Todd[a], and David E. J. Armstrong[a]

*a) Department of Materials, University of Oxford*

Parks Road, Oxford, OX1 3PH, UK

email: *alex.leide@bristol.ac.uk*

*Abstract.* Raman spectroscopy has been used to identify defective bonding in single crystal 6H-SiC irradiated with neon and silicon ions. Different defects are observed with neon and silicon ion implantations, suggesting that the choice of ion is important in radiation damage studies. Raman spectra of ion irradiated SiC show less tensile strain on the Si-C bonds than neutron irradiations in the literature, explained by a residual compressive stress caused by radiation swelling constrained by the undamaged substrate. Evidence of oxidation during high temperature ion implantation is observed as C-O and Si-O Raman signals. Annealing irradiated SiC while acquiring Raman spectra shows rapid recovery of Si-C bonding, but not a complete recovery of the unirradiated structure. Annealing irradiated SiC causes surface oxidation whereas unirradiated SiC does not oxidise. Comparisons are made to the apparent radiation resistance of diamond and silicon which have similar crystal structures, but consist of a single element, leading to the suggestion that chemical defects are responsible for increased radiation damage in SiC.

Keywords: Raman spectroscopy; radiation defects; ion irradiation; residual stress; silicon carbide

---

[1] Now at School of Physics, University of Bristol, Tyndall Avenue BS8 1TL

# 1  Introduction

Silicon carbide (SiC) is a structural ceramic material useful in extreme environments, primarily for aerospace and nuclear applications, because of its excellent high-temperature properties, including creep resistance, high strength at elevated temperatures, corrosion resistance and general chemical inertness, high thermal conductivity, and low thermal expansion coefficient [1–3]. SiC is desirable for applications as fission fuel cladding, or as a component of the blanket of fusion reactors due to its low neutron absorption cross-section, low level of long-lived radioisotopes, and stability under high temperature-high dose neutron irradiation [1,4–9].

For nuclear applications, a thorough understanding of radiation defects and their effects on material properties is required to evaluate the suitability of a material for its application, and to predict the evolution of its properties over time. To accelerate radiation damage processes and material investigations, ion implantation is commonly used as a surrogate for neutron irradiation [10,11]. It allows displacement damage to be introduced to a material in controlled conditions in a matter of hours as compared to many days for comparable damage in a nuclear fission reactor. Additionally, it does not introduce radiological hazards due to sample activation, avoiding the requirements for specialist "active" laboratories, sample cooling, and remote handling. While ion implantation is somewhat able to replicate certain aspects of neutron irradiation, questions remain about the defect structures produced by ion implantation, their equivalence to those from neutron irradiation, and the effect these may have on material properties.

Raman spectroscopy has been used to identify defective bonding in damaged SiC with early research concentrating on elucidating different amorphous structures of SiC [12–15], and processing induced defects [16]. Only more recently has Raman spectroscopy been applied to radiation defects in SiC in the context of nuclear reactor applications. The power of this technique is that chemical defects invisible in the TEM and diffraction experiments are identifiable, which helps to understand radiation damage mechanisms [17]. Additionally, Raman spectroscopy can be applied to relatively large volumes of bulk specimens compared to TEM, avoiding residual stress relief during TEM specimen production, and the effects of free surfaces in small sample volumes which can reduce the measured defects through recombination [18]. Similar chemical information to Raman spectroscopy can be obtained using electron energy loss spectroscopy in a STEM. Typical interaction volumes in a STEM are approximately $10^6$ times smaller than in confocal Raman spectroscopy which has an interaction volume approximately 1 µm$^3$.

Typical features of Raman spectra after irradiation are reduction in intensity and broadening of the characteristic Si-C peaks and the appearance of peaks corresponding to Si-Si and C-C bonding. These are apparent in all irradiation conditions and will be discussed in more detail below. Sorieul *et al.* and Chaâbane *et al.* summarised the peaks observed in irradiated SiC based on their observations and earlier literature (Table 1) and this has widely been used in the more recent literature for signal identification [19,20]. They cite Bolse *et al.* [12] for the 660 cm$^{-1}$ peak of distorted SiC, however this paper makes no reference to a peak at 660 cm$^{-1}$ . A further literature search for appropriate identification of this peak found that it can be attributed to C-O-C vibrations in solid carbon dioxide with a tetrahedral structure [21].

*Table 1: Identification of Raman signals in SiC from Sorieul et al. and Chaâbane et al. [19,20]. Si-C peaks in parentheses are second order Raman peaks with low intensity. The peak marked with an asterisk at 660 cm$^{-1}$ attributed to distorted SiC is discussed in the text.*

|  | Si-Si (cm$^{-1}$) | Si-C (cm$^{-1}$) | C-C (cm$^{-1}$) |
|---|---|---|---|
| **Unirradiated** |  | (146)    767 <br> (206)    789 <br> (505)    966 |  |
| **Irradiated** | ~188 <br> ~260 <br> ~535 | 766, 789, 966 <br> ~600, ~660* Distorted SiC <br> ~870 Highly disordered SiC <br> ~933 Slightly disordered SiC | ~1080 sp$^3$ <br> ~1420 mixed sp$^2$/sp$^3$ D band <br> ~1600 sp$^2$ G band |

Raman spectra of neutron irradiated SiC are presented in references [17,22–25] for a range of temperatures and doses in both single crystal 6H-SiC and polycrystalline 3C-SiC. Damage appears to reduce with increasing temperature until at 1180 °C the Raman spectrum after a dose of 1.9 dpa maintains sharp peaks similar to the unirradiated material [17]. Increasing dose at a given temperature causes a lower Si-C peak intensity until a saturation of apparent damage [17]. Initially a D band C-C signal is visible after 2 dpa at 600 °C, with the addition of a G band signal after 12 dpa [25]. The D/G ratio can indicate the extent of carbon ordering in the structure – higher G band intensity corresponds to extended ordered sp$^2$ carbon structures similar to graphite (graphene), although not necessarily ring shaped, while the D band is indicative of disordered C-C pairs or small chains [26].

Both the TO and LO Si-C peaks are shifted to lower wavenumbers with increasing neutron dose – this shift being reduced with increasing temperature [17]. The shift to lower wavenumbers is indicative of tensile strain on the Si-C bonds, in agreement with the observed swelling characteristics of irradiated SiC [27].

Raman spectroscopy has commonly been used with ion implanted samples as the probed depth of confocal Raman microscopy is similar to the damaged depth of ion implantation [19,20,28–30]. The paper by Chen *et al.* is the only direct comparison of neutron irradiations matched with similar nominal dpa ion irradiations [28]. They chose silicon and carbon ions to reduce chemical effects, but based on their Raman spectra these ions may have exacerbated the chemical effects due to the propensity of Si and C to form covalent bonds with themselves and each other. The general form of the ion irradiated spectra matches the neutron irradiated spectra with the broadening of Si-C peaks and appearance of Si-Si and C-C peaks. However, two significant differences exist: 1. Si-C peaks are not shifted to as low wavenumbers after ion irradiation as after neutron irradiation, indicating less tensile strain on Si-C bonds. 2. carbon ion irradiation produces more intense D and G band signals than neutron irradiation, while Si ion irradiation only produces a D band signal. There appears to be no difference to the Si-Si bonding. This work demonstrates significant issues in the choice of ions for ion implantation if the aim is to directly replicate neutron irradiations.

Dual beam ion implantations have shown a synergistic effect of displacement damage with helium ions stabilising different defects compared to silicon ions alone [29]. Carbon-carbon defects appeared more strongly in Raman spectra with helium implanted alongside silicon ions than with silicon ions alone. This is attributed to helium-vacancy complexes and stabilisation of extended defect clusters preventing carbon recombining with nearby vacancies, leading to the stronger C-C signal.

The existing literature on Raman spectroscopy characterisation of radiation damage indicates that ion implantation conditions can be tailored to closely replicate neutron damage in terms of Si-Si and Si-C defects, but the C-C defects appear difficult to replicate, as does the shift to lower wavenumbers of the Si-C peaks caused by tensile swelling strain [19,20,28–30].

Since Raman spectroscopy, as used in this context, is providing information about interatomic bonding, it is convenient to describe the structure of SiC in terms of interatomic bonds rather than crystal symmetry.

An alternative to symmetry-based approaches to structure is to use topology based approaches where the crystal is considered a network of bonds connecting nodes (atoms) [31]. The crystallographic unit cells shown in Figure 1 (a) and (b) for 6H- and 3C-SiC appear vastly different and respond differently to diffraction experiments. However, topologically in terms of their interatomic bonds, 6H- and 3C-SiC are very similar (Figure 1 (c) and (d)) and one would expect only small differences in terms of radiation defects. Snead *et al.* showed that under low temperature (~60 °C) neutron irradiation when defect annealing is minimal, single crystal 6H- and 3C-SiC behave identically to nanocrystalline CVD 3C-SiC in terms of swelling, point defects, and amorphisation [32]. Only at higher temperatures when defect annealing becomes apparent does the "nuclear grade" nanocrystalline CVD β-SiC become radiation tolerant. In the same work, single crystal silicon showed no change in lattice parameter or bulk swelling [32].

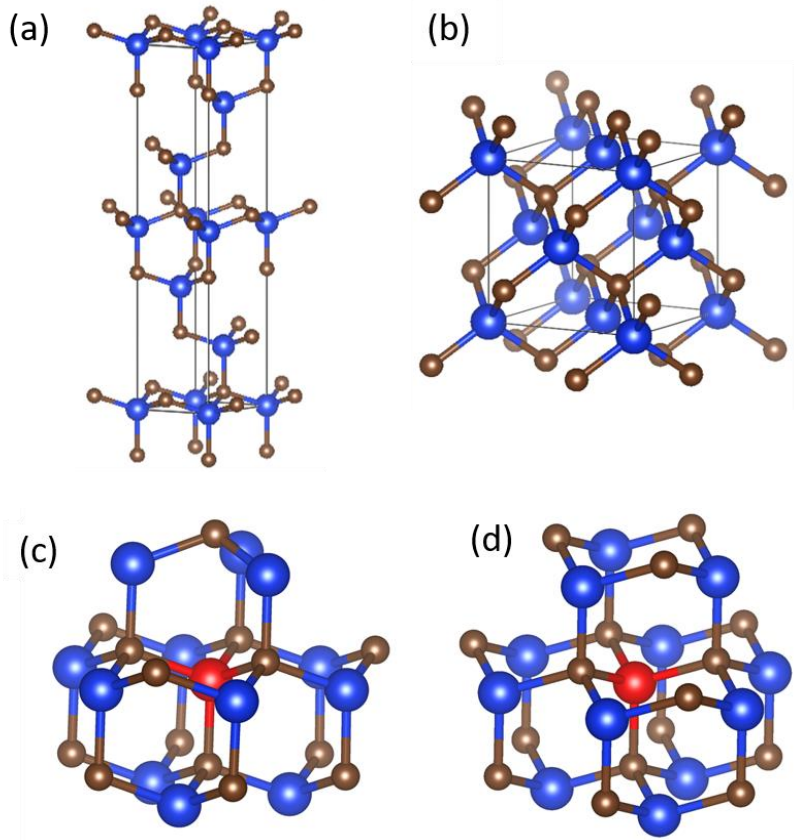

*Figure 1: Crystallographic representations of atomic structures where brown circles are carbon atoms, and blue are silicon atoms. (a) unit cell of 6H-SiC showing stacking sequence. (b) unit cell of 3C-SiC. (c) and (d) are local cluster topological representations of α-SiC and β-SiC centred on the central red silicon atom. This representation demonstrates how α- and β-SiC are structurally very similar, despite the stark differences in the crystallographic representations, and the resulting differences in diffraction-based experiments.*

Diamond, silicon, and silicon carbide are connected as {4,4} tetrahedra, and structurally, 3C-SiC is identical to silicon and diamond but with different lattice parameters and occupation of lattice sites [31]. This is particularly important for silicon carbide where four types of bond are possible with different covalent bond lengths which must be accommodated in a defective network: Si-C (1.89 Å), Si-Si (2.35 Å), C-C ($sp^3$ 1.54 Å, $sp^2$ 1.47 Å). Additionally, point defects in SiC have a net positive volume due to strains caused by changes in bond lengths [33]. Only the $C_{Si}$ anti-site has a negative volume associated with it. This net positive volume of defects in SiC contributes to volumetric swelling under irradiation [27], resulting in residual stresses in ion irradiated samples where swelling is constrained by a substrate [34].

While the general properties of radiation defects in SiC have previously been studied, some specific details are explored here. This work aims to investigate the bonding defects in ion irradiated SiC, specifically the effects of constrained swelling, and the chemical effect of implanting a self-ion compared to a noble ion. The role of temperature in formation of radiation defects and defect annealing is explored, alongside radiation-enhanced oxidation. These atomic level defects are related to macroscopic effects such as swelling and mechanical properties. Diamond and silicon are used as monatomic references where the effect of chemical defects should be removed.

## 2 Methods

A pre polished sample of 6H-SiC single crystal, with surfaces parallel to the (0001) basal plane was purchased from Pi-Kem Ltd (Tamworth, UK). Ion implantation was carried out at the Surrey Ion Beam Centre, UK using the 2 MV Van de Graaf accelerator. Samples were gently clamped to a heated stage using washers to blank part of the specimen from the ion beam. The stage was held at 300 °C (±5 °C) or 750 °C (±10 °C) in a vacuum of ~$1\times10^{-6}$ mbar. The samples were implanted with neon ions at three energies (1450 keV, 720 keV, and 350 keV) or silicon ions at three energies (2000 keV, 1000 keV, 500 keV) to create a flattened damage profile within the interaction volume of confocal Raman spectroscopy (Figure 2). Neon ions were chosen to investigate any potential chemical effects from implanted silicon ions, while producing a similar displacement damage profile. The number of displacements per atom (dpa) was calculated using the Stopping and Range of Ions in Matter (SRIM) Monte Carlo code with the quick Kinchin-Pease model [35,36]. Displacement energies for silicon and carbon were 35 eV and 21 eV respectively, with binding energies set to 0 eV [37]. Target density was set to 3.21 g/cm$^3$. The peak nominal damage is ~2.5 dpa. A low dose silicon implantation was conducted at 750 °C with a nominal peak damage of ~0.25 dpa.

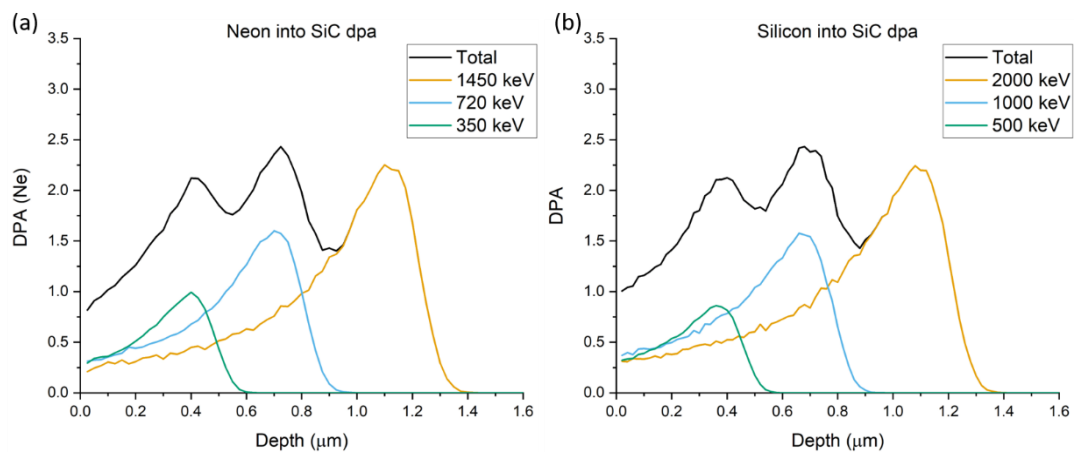

*Figure 2: Damage profile of (a) neon and (b) silicon ion implantation into 6H-SiC, as calculated from SRIM* [35]

Raman spectra were acquired using the Witec Alpha 300R confocal Raman microscope in the Materials Research Facility at UKAEA. A green 532 nm laser operating at 10 mW was used with 5 accumulations of 0.2s each. Light was collected using a ×100 objective lens and sent to the spectrometer via a 50 µm diameter optical fibre which acted as the confocal aperture giving depth resolution of approximately 1 µm [38]. A diffraction grating with 1800 l/mm was used to scatter light onto the CCD.

For the annealing experiment a Linkam TS1500 environmental stage was fitted to the Raman microscope, allowing heating up to 1500 °C at 200 °C/min. The sample was placed in an alumina crucible on the heating stage with a quartz window. Light was collected through a ×50 long focal length objective lens for 60s approximately every 2 minutes during the 1000 °C annealing experiment. Long exposures were required as the long focal length collection optics are not as efficient as the ×100 objective lens which would normally be used.

Additionally, Raman spectra were collected from single crystal silicon (100) which had been irradiated in the same conditions as described above, and a sample of natural diamond which had been implanted with five energies of lithium ions at 500 °C to create a flattened damage profile over a depth of 300 nm, peaking at 1.7 dpa. These samples will be discussed as comparisons to the silicon carbide, and in the context of our other work on nanoindentation of irradiated 6H-SiC [34].

# 3 Results

## 3.1 Raman spectra

Figure 3 shows Raman spectra of 6H-SiC single crystals before and after ion irradiation. The unirradiated spectrum is taken from a blanked region of the sample irradiated at 750 °C and the intensity is scaled by ×0.25. All the irradiated spectra show significant damage; reduction in intensity, peak asymmetry and broadening, and the appearance of extra peaks for Si-Si and C-C bonding. This makes exact peak positions and integrated areas difficult to identify. It is worth pointing out that this type of damaged spectrum does *not* indicate amorphisation despite the significant changes. TO and LO Si-C peaks remain distinguishable, the first order Si-Si peak remains narrow and at a higher wavenumber than in amorphous SiC, and there is no strong signal from amorphous C-C bonds [20]. Additionally, the swelling observed in these samples is too small for amorphisation and the corresponding 10% volume expansion to have occurred [34].

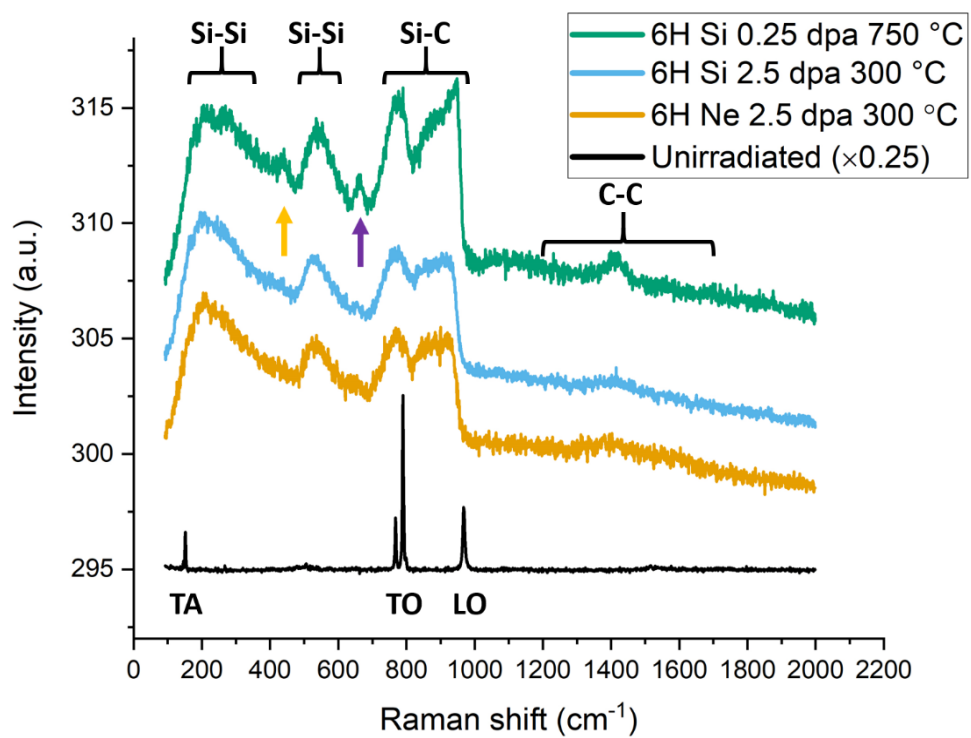

*Figure 3: Raman spectra of unirradiated and ion implanted single crystal 6H-SiC. The unirradiated spectrum is multiplied by ×0.25 in the y axis and is from the unirradiated region of the 750 °C irradiated specimen. Orange arrow points at a peak for Si-O-Si bonding, and the purple arrow to C-O-C bonding.*

The peak at ~530 cm$^{-1}$ corresponds to Si-Si bonding. The 535 cm$^{-1}$ Si-Si peak in the 0.25 dpa 750 °C spectrum is 1.4× taller relative to the background than the 2.5 dpa 300 °C spectra. The broad signal at lower frequencies, < 400 cm$^{-1}$ is a second order peak of silicon and is not used for analysis. The peaks associated with Si-C bonding remain sharper and with higher intensity after 0.25 dpa 750 °C Si ion implantation than the 300 °C 2.5 dpa implantations. The 300 °C 2.5 dpa spectra are similar in the Si-C region regardless of whether neon or silicon ions were implanted.

The sample irradiated at 750 °C contains additional signals labelled with orange and purple arrows in Figure 3. The peak at ~660 cm$^{-1}$ (purple arrow) is associated with C-O-C bonding as found in quartz-like $CO_2$ [21]. The small shoulder at ~430 cm$^{-1}$ could be due to frequencies associated with Si-O-Si bonding, as in α-cristobalite or other silica polymorphs [39,40]. This is difficult to confirm due to the weak signal and because the other characteristic silica peaks are around 180-300 cm$^{-1}$ and are obscured by the broad Si-Si signal. These oxide signals are not clearly visible in the samples irradiated at 300 °C. No surface oxidation was observed microscopically. The unirradiated spectrum also shows no signals corresponding to Si-O and C-O bonds despite being in the same environment at 750 °C.

All three irradiation conditions show signals corresponding to C-C bonding, but in different ways (Figure 4). The samples implanted with silicon ions have a single peak at ~1410 cm$^{-1}$ corresponding to D band sp$^2$ type C-C bonding, most likely in isolated pairs or short chains [24]. The higher temperature lower dose sample (green line in Figure 4) has a taller peak suggesting more D band sp$^2$ type C-C bonding than the high dose low temperature (blue line). The neon irradiated sample has two overlapping peaks with contributing signals from both the D and G bands of sp$^2$ carbon at ~1400 cm$^{-1}$ and ~1580 cm$^{-1}$. This is associated with graphitic clusters rather than isolated pairs or chains of carbon atoms in the network [24].

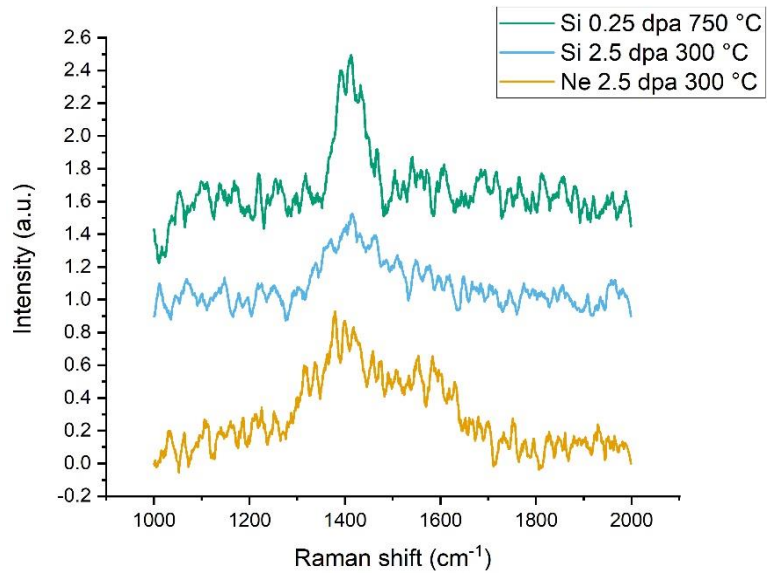

*Figure 4: Spectra from the C-C bonding region of irradiated 6H-SiC with a straight baseline subtraction. The silicon irradiated samples have a single peak at ~1410 cm$^{-1}$ while the neon irradiated sample has a double peak at 1400 cm$^{-1}$ and 1580 cm$^{-1}$.*

## 3.2 In-situ annealing of irradiated SiC

A sample of 6H-SiC irradiated with neon ions at 300 °C was annealed at 1000 °C in air on a heated stage in the Raman microscope, and spectra were acquired during this process, shown in Figure 5. The spectra acquired at 1000 °C are shifted to lower wavenumbers than spectra acquired at room temperature, corresponding to thermal expansion of the sample, which is the basis of Raman thermometry [41,42]. Peaks are shifted by different amounts as different bond resonances are affected differently by temperature.

After a very short time, characteristic SiC peaks begin to form, then sharpen with time. Almost all changes occurred in the first 20 minutes. Spectra were acquired for another hour (line 10) to ensure no more structural evolution is occurring, before cooling to room temperature in ~10 minutes. Stage drift during the hour between line 9 and 10 likely explains the reduction in intensity by moving the specimen out of focus.

Initially the Si-Si peak increases in intensity and narrows (lines 1-3), then begins to disappear with longer times (line 6). Initially there are two C-C peaks at ~1350 and 1480 $cm^{-1}$ (line 2). The 1350 $cm^{-1}$ peak disappears and the 1480 $cm^{-1}$ peak increases in relative intensity (line 4). A C-O-C peak appears clearly at ~650 $cm^{-1}$ (lines 2, 3, and 4), then disappears by line 6. The spectrometer diffraction grating did not allow collection of the low wavenumber end of the spectrum at the same time as the high wavenumber carbon end so the evolution of Si-O (~430 $cm^{-1}$) bonding cannot be seen.

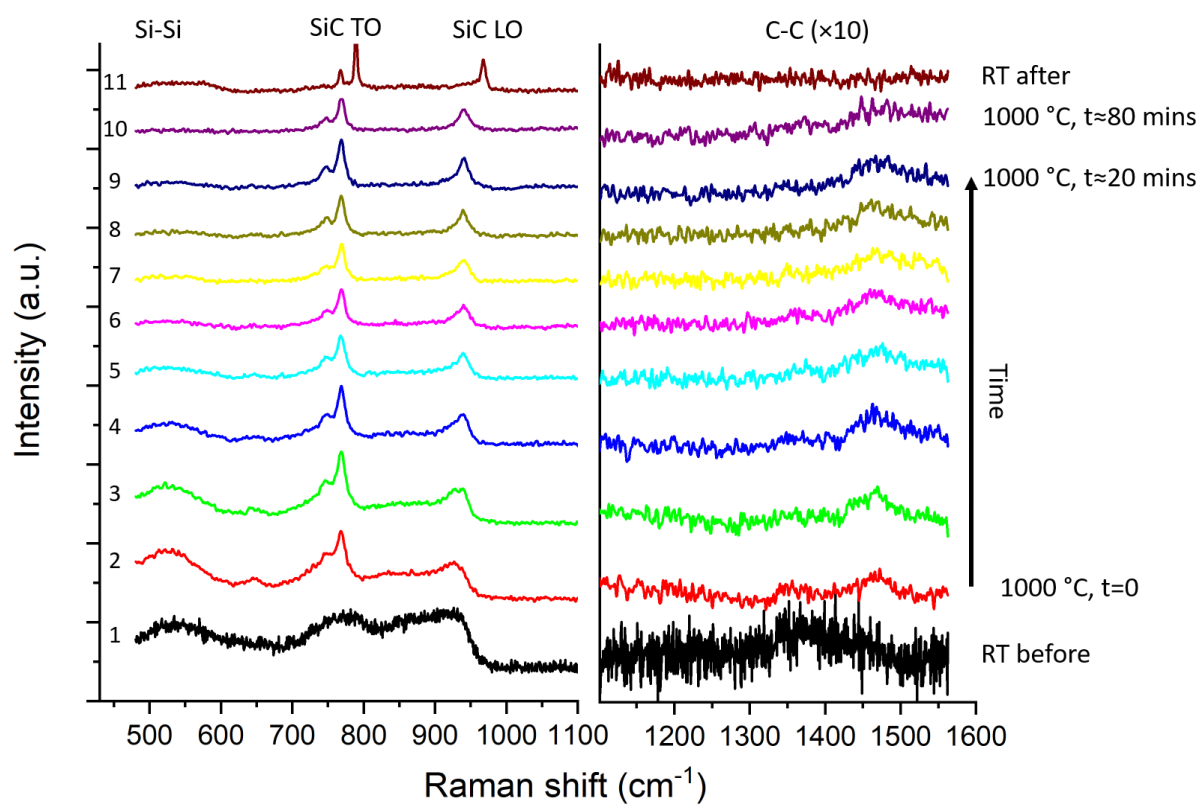

*Figure 5: Raman spectra of 6H-SiC acquired during annealing at 1000 °C in air. "Before" and "After" lines were measured at room temperature using the same ×50 long focal length objective and heating stage window as the high temperature spectra. The C-C region of the spectra are magnified ×10 to see these low intensity peaks more clearly.*

As received, the pristine 6H-SiC single crystals were translucent green. Irradiation changed them to opaque grey on the irradiated surface. After annealing, the sample returned to translucent green, but not entirely to the pristine state. This is apparent when the "after" Raman spectrum (line 11 in Figure 5) is compared to the unirradiated spectrum shown in Figure 3; there is still a small, broad Si-Si signal and asymmetry to the SiC LO signal, changing the optical properties.

Optical micrographs showed sub-micron features on the surface of the annealed specimen, some of which were in larger clusters with a featureless ring around them (Figure 6 (a)) These features remained on the sample after cleaning with ethanol and isopropanol. Closer inspection with SEM-EDX (Figure 6 (b) and (c)) revealed that these particles are silica and that the irradiated SiC had oxidised during annealing. The small oxygen peak in the "background" EDX line in Figure 6 may be due to the thin native oxide layer on SiC, or oxygen possibly incorporated into the SiC bulk during irradiation. A sample of unirradiated 6H-SiC which had been annealed in air on the high temperature stage in the same Raman microscope showed no signs of surface oxidation.

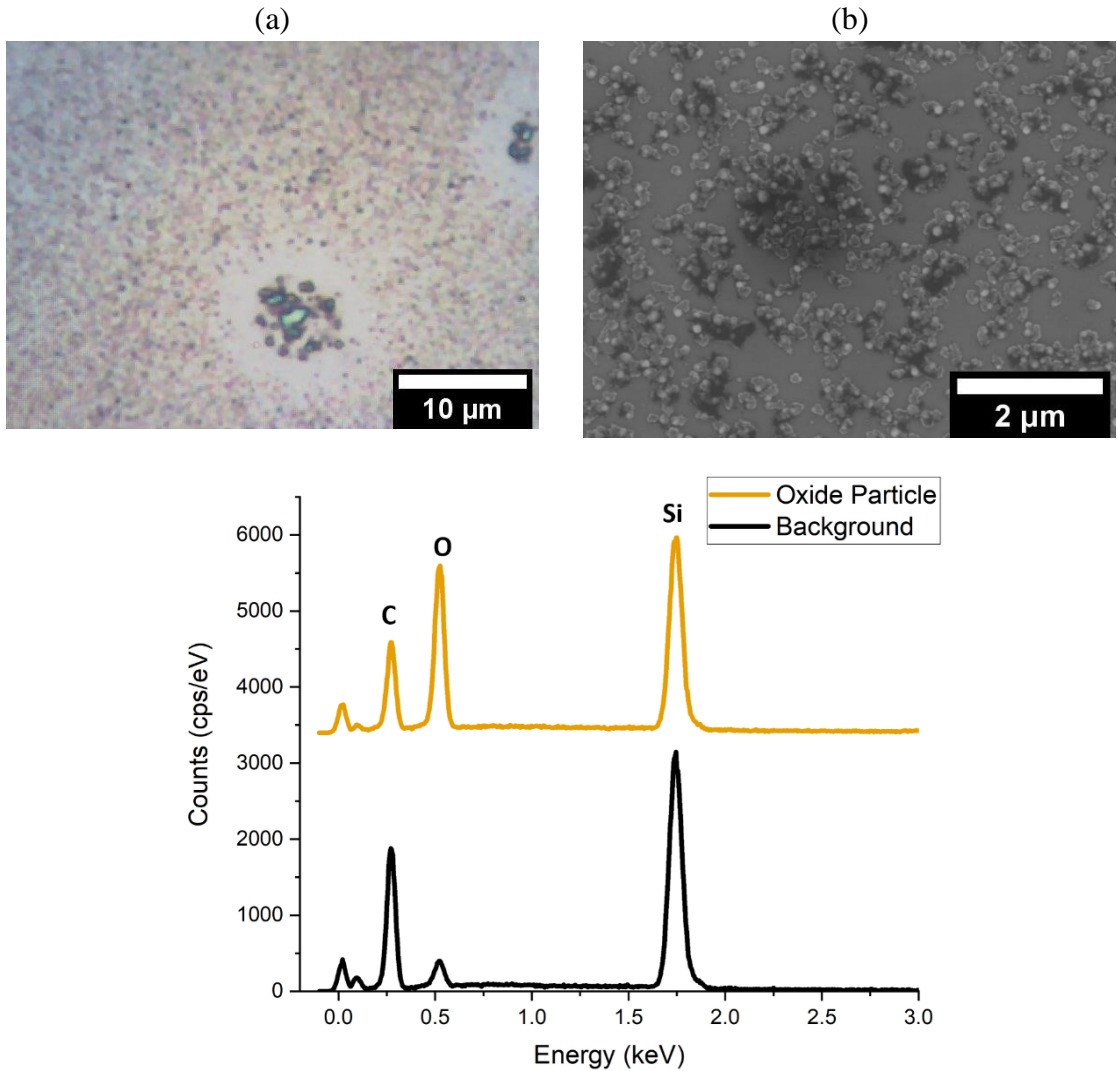

*Figure 6: (a) Optical and (b) Secondary electron images of neon irradiated 6H-SiC after annealing at 1000 °C in air. Small sub-micron oxide particles are visible, some of which have clustered into larger particles. Below are EDX spectra of the background material and particle*

## 3.3 Single element crystals

Raman spectra of ion implanted single crystal silicon and natural diamond show no change in peak position (Figure 7), remaining at the unstressed, unirradiated positions of 520 cm$^{-1}$ and 1332 cm$^{-1}$ respectively. There was no sign of graphitisation of the diamond. The reduction in intensity may be a true reflection of damage to the crystal lattice, or it may be due to the optical focus changing.

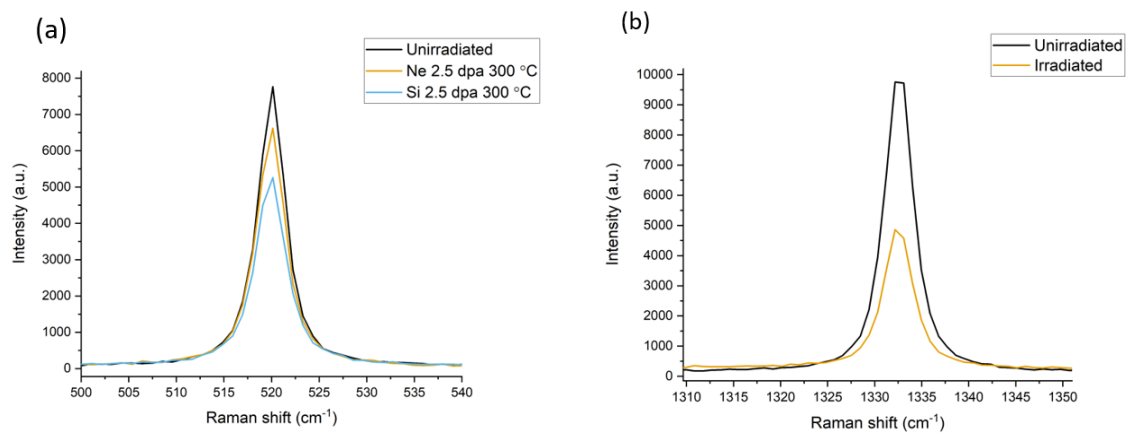

*Figure 7: Raman spectra of (a) irradiated silicon single crystals, and (b) irradiated natural diamond. Spectrometer and laser settings were kept constant; however, the optical focus may have changed while moving between samples contributing to the change in intensity.*

# 4 Discussion
## 4.1 Choice of implanted ion species and chemical defects

The Si-Si and Si-C regions of the Raman spectra of irradiated silicon carbide are similar regardless of whether silicon or neon ions are used. The significant difference is in the C-C region where different carbon bonding is observed (Figure 4). Neon, being an inert gas, will not contribute to covalent bonding in the lattice, while silicon ions can substitute into a silicon carbide lattice via covalent bonding to both silicon and carbon atoms. These additional silicon ions will prevent any extended carbon chains or rings from forming during atomic displacements, precluding the observation of a G band in the Raman spectrum. Isolated pairs or short chains of carbon atoms remain, so the D band is present. In neutron irradiations, there is no implanted ion which could break up carbon networks, so these can evolve to contribute to a G band signal. Literature using carbon ions shows large D and G band signals, as well as $sp^3$ carbon which are not seen in neutron irradiations nor other ion irradiations [19,28]. This reinforces the evidence that silicon and carbon ions should be avoided for ion implantations of SiC due to their chemical effect on defects. In metals self-ions are often favoured as they avoid compositional changes, however in multi-element ceramics it seems that self-ions may be the worst choice of ion.

Chemical defects (in particular C-C homonuclear bonds) in silicon carbide have been shown as a vital step towards amorphisation, and are present in radiation damaged SiC prior to amorphisation [20,43–45]. Silicon and diamond are not able to form chemical defects; atomic displacements in silicon will simply form Si-Si bonds, and in diamond $sp^3$ C-C bonds. Radiation induced defects in silicon are mobile even at room temperature, and by 300 °C most defects will be thermally annihilated [46]. Although graphite is the thermodynamically stable phase of carbon, defects in a diamond lattice are likely to favour recombination into metastable diamond as long-range stochastic displacements and a large volume expansion would be required to transform into graphite. The constraint of a stiff diamond network makes this kind of transformation very difficult. No phase change towards graphite is observed in the Raman spectra in Figure 7. Single element crystals appear to be quite resistant to radiation defects, and the nanoindentation properties of silicon and diamond show no change after ion irradiation [47].

## 4.2 Structural damage and crystal strains

Broad peaks in the Raman spectra indicate a range of bonding environments with variations in bond length and angle as defects attempt to fit into the damaged crystal lattice. Despite this damage, the sample is still crystalline, with Raman spectra not showing typical features of amorphous SiC. Electron backscatter diffraction was performed on the unimplanted and implanted surfaces of these specimens as part of our work on residual stress [34]. Electron backscatter diffraction patterns could be formed in the implanted region; however, these were noticeably blurred and diffuse compared to patterns formed in the unirradiated region (Figure 8). This suggests some degradation of the crystal lattice and changes to the structure factor of the irradiated crystal. These diffraction patterns originate from the top ~40 nm of the specimen where damage is lower (~0.9 dpa) than in the volume contributing to the Raman signal (Figure 2).

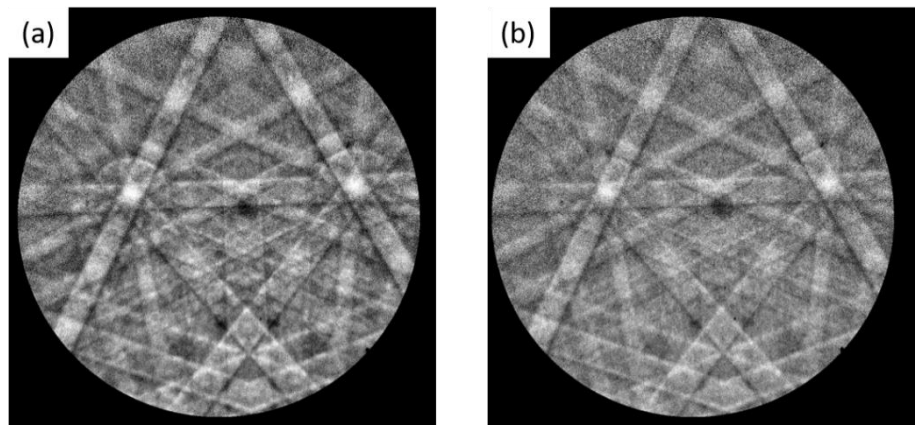

*Figure 8: EBSD patterns from neon irradiated 6H-SiC. (a) is from the unimplanted region, (b) is from the implanted region. Both patterns were acquired with identical microscope settings in the same mapping scan.*

The four types of covalent bond which can be formed in defective SiC have different equilibrium bond lengths which must distort to maintain connectivity in the lattice. The broad Si-Si peak is at ~530 cm$^{-1}$ in Figure 3, however in crystalline silicon the first order Raman peak position is at 520 cm$^{-1}$ [48]. This shift to a higher wavenumber indicates compressive strain on the Si-Si bonds, i.e. they are shorter than their equilibrium length to fit into the damaged silicon carbide lattice which consists mostly of shorter Si-C bonds. Similarly, the radiation damaged Si-C peak position is shifted to a lower wavenumber than in the perfect crystal, indicating a longer bond length. However, in neutron irradiated SiC, Raman spectra show a larger shift to lower wavenumbers as the specimen is allowed to swell in all directions and the Si-C bond length is longer than in ion implanted SiC [17,22,23,28]. This is a consequence of the lateral constraint in ion irradiation experiments, resulting in biaxial compressive stress of several GPa [34]. Swelling is allowed out of the free surface; thus, the small elongation of Si-C bonds may be attributed to swelling in this direction.

It is well known that higher irradiation temperatures encourage more defect annealing in SiC and a smaller degradation of properties such as volumetric swelling, thermal conductivity, and radiation-induced hardening [47,49,50]. This is seen in the sharper peaks of the Raman spectrum irradiated at 750 °C. At 750 °C silicon defects are able to recombine in addition to carbon defects, whereas at 300 °C only carbon defects can recombine, leading to less residual damage to the SiC lattice at high temperature. This highlights the importance of using SiC at high temperatures in radiation environments. Our other work in polycrystalline reaction-bonded SiC shows that damage saturates below 0.25 dpa at 750 °C, so 0.25 dpa in single crystal 6H-SiC should be comparable to 2.5 dpa at 750 °C [47].

Homonuclear Si-Si and C-C bonds cause significant strains to the SiC lattice which contribute to swelling and ultimately amorphisation [33,51]. As swelling strains are related directly to structural defects, the presence of residual biaxial compressive stresses (or prevention of swelling), is likely to influence the defects which are able to form, hindering defects with a larger positive volume [33]. Strain gradients in the z (out of surface) direction in ion irradiated SiC have been shown to influence defect formation, diffusion, and recombination [52–54]. The preference for D band C-C defects seen in Raman spectra of ion irradiated SiC may be due to their smaller size than G band C-C defects which requires a larger C-C network closer in structure to graphite rings. Conrad *et al.* suggest that residual compressive stresses on covalent bonds may encourage amorphisation to relax the volume and the excess energy of the constrained bonds [13]. Li has shown that large interatomic strains due to constrained homonuclear bonds are relieved by amorphisation and geometric relaxation [51]. This may explain the higher critical amorphisation temperature for ion implantations than neutron irradiations [55]. The undamaged substrate is influencing amorphisation through additional constraint. By extension, an undamaged substrate would be expected to influence defects in a damaged layer.

## 4.3 Effect of radiation damage on oxidation

Evidence of Si-O and C-O bonding is observed in the Raman spectra of SiC irradiated at 750 °C, despite no surface oxidation being visible in optical microscopy or SEM; the Si-O and C-O signals are coming from within the Raman laser interaction volume which corresponds to the ion implanted layer. Tetrahedral C-O-C or Si-O-Si bonds should substitute satisfactorily into a damaged tetrahedral SiC network [31]. Although the ion implantation stage was held in a good vacuum ($\sim 1 \times 10^{-6}$ mbar) some oxygen inevitably remained in the chamber. Unirradiated regions of the same specimens did not have Si-O or C-O signals which suggests the oxidation is enhanced or caused by radiation defects as the clamping washer is unlikely to make a gas-tight seal. This is supported by the annealing experiments whereby samples irradiated at 300 °C annealed at 1000 °C oxidised, and unirradiated annealed samples did not. Oxide formed during ion implantation could have been mixed into the bulk below the surface by ion beam mixing and the directionality of displacement damage during ion implantation.

Oxidation during the annealing experiment resulted in silica particles on the sample surface, unlike the oxidation caused during high temperature ion irradiation, suggesting a different mechanism. Si-O bonds formed during annealing post irradiation are not mixed into the damaged ion implanted layer allowing them to develop into microscopic particles rather than being incorporated into the damaged SiC structure. C-O bonds created during annealing might be expected to evaporate as $CO_2$ molecules.

Enhanced oxidation of defected SiC is not unreasonable. Molecular dynamics simulations have shown that incoherent grain boundaries oxidise faster than coherent grain boundaries and perfect crystalline SiC surfaces [56]. This is due to undercoordinated bonds on the incoherent grain boundaries which enhanced their reactivity with oxygen. Radiation damage also produces undercoordinated bonds in SiC which will enhance oxidation. McHargue and Williams found that ion beam amorphised SiC etched 2.5-4 times faster than un-irradiated samples in a boiling solution of potassium ferricyanide and potassium hydroxide, and had an oxide layer 1.7 – 3 times as thick after 1 hour in flowing oxygen at 1300 °C [57]. This enhanced oxidation or corrosion would be expected to occur in both ion implanted and neutron irradiated SiC and is likely to be an issue if oxygen concentrations are not carefully controlled. Experiments with different steam chemistries and pressure have shown that SiC can form a passive oxide layer resistant to erosion in flowing coolants [58,59]. However, this passive layer may be affected by radiation degradation  It is more concerning for $SiC_f$/SiC composites where the interphase layer is especially susceptible to oxidation degradation [58].

The presence of undercoordinated atoms at (or near) irradiated surfaces is likely to enhance corrosion which may be important for the application of SiC in other corrosive media including fusion-relevant lithium-containing tritium breeding materials. This does not appear to have been investigated significantly with all relevant corrosion tests being conducted on unirradiated specimens. Pores respond poorly to corrosion in Pb-Li, but a dense CVD SiC coating provided much improved protection [60,61]. Ion implantation may be a suitable surface modification mechanism for creating radiation damage relevant for studying radiation-enhanced corrosion of SiC without the issues of sample activation from neutron irradiation.

The in-situ annealing experiment demonstrates the rapid recovery of radiation damage at 1000 °C. This has implications for high temperature examination and testing of irradiated SiC (and likely other ceramic materials) as damage may be recovered before the measurement is made leading to incorrect conclusions.

# 5    Conclusions

Raman spectroscopy is a powerful technique to investigate radiation damage which is invisible in a TEM. In particular it allows investigation of chemical defects which are important in silicon carbide due to the variety of bond types which are possible between silicon and carbon atoms, and the effects these have on distorting the lattice. Silicon and carbon ions should be avoided for implantation into SiC as they alter defect chemistry, possibly creating defect structures which are unrepresentative of neutron irradiation. Inert ions should be chosen for radiation damage studies so as not to influence the defect chemistry and covalent nature of the lattice.

In contrast, single atomic species tetrahedral ceramics appear to be remarkably radiation resistant compared to SiC at the temperatures in this work, as defects simply result in the same type of bonds reforming, and no significant lattice distortion and internal stresses. This would suggest that SiC is *not* radiation resistant inherently, indeed it is particularly susceptible to radiation damage due to chemical defects. It appears resistant in nanocrystalline form due to a high density of defect sinks which are particularly effective at elevated temperatures [62]. Structurally, α- and β-SiC are very similar, and their defects are also similar, so any radiation resistance of β-SiC is likely to be due to microstructural design.

As well as information on chemical defects, Raman spectroscopy identified that ion implanted SiC does not experience as large a tensile strain as neutron irradiated SiC in the literature. This is likely to be due to the constraining effect of the undamaged substrate, which causes residual compressive stresses and significant changes in mechanical properties [34]. This compression also has an influence on defect structures; it is not simply a lattice strain in isolation. Defects caused by ion irradiation are identifiably different to neutron irradiation defects.

Radiation defects appear important in corrosion and oxidation of SiC and should be considered when studying corrosion of materials for nuclear applications.

Although ion implantation appears not to recreate neutron irradiation defects in SiC, it is still useful for introducing some displacement damage, as long as these limitations are considered in analysis of results. The displacement damage and associated measurements may not be representative of materials subjected to "real" neutron irradiation conditions.

# 6 Acknowledgements

Thanks to Dr Nianhua Peng at Surrey Ion Beam Centre for running the ion implantation, and to Chris Smith at the Materials Research EurofusionFacility, UKAEA, for training on the Raman microscope. Thanks to Dr Bo-Shiuan Li, Prof. Steve Roberts and Prof. Sir Peter Hirsch for providing the irradiated diamond sample. Raman spectroscopy was carried out at UKAEA's Materials Research Facility which has been funded by and is part of the UK's National Nuclear User Facility and Henry Royce Institute for Advanced Materials. The authors acknowledge use of characterisation facilities within the David Cockayne Centre for Electron Microscopy, Department of Materials, University of Oxford, alongside financial support provided by the Henry Royce Institute (Grant ref EP/R010145/1). Financial support through the EPSRC Science and Technology of Fusion CDT [grant number EP/L01663X/1] is gratefully acknowledged. This work has been carried out within the framework of the EUROfusion Consortium and has received funding from the Euratom research and training programme 2014-2018 under grant agreement No 633053. The views and opinions expressed herein do not necessarily reflect those of the European Commission.